# Ferroptosis as a Biological Phase Transition I: avascular and vascular tumor growth

J.M. Nieto-Villar [1][*] & R. Mansilla [2][*]

[1] Department of Chemical-Physics, A. Alzola Group of Thermodynamics of Complex Systems of M.V. Lomonosov Chair, Faculty of Chemistry, University of Havana, Cuba. [2] Centro de Investigaciones Interdisciplinarias en Ciencias y Humanidades, UNAM, México. * Corresponding author. Email address: nieto@fq.uh.cu (J.M. Nieto-Villar), mansy@unam.mx (R. Mansilla)

**Abstract**

Herewith we discuss a network model of the ferroptosis avascular and vascular tumor growth based on our previous proposed framework. Chiefly, ferroptosis should be viewed as a "first order" phase transition characterized by a supercritical Andronov–Hopf bifurcation, with the emergence of limit cycle. The increase of the population of the oxidized PUFA fragments, take as the control parameter, involves an inverse Feigenbaum, (a cascade of saddle-foci Shilnikov's bifurcations) scenario, which results in the stabilization of the dynamics and in a decrease of complexity.

## 1. Introduction

According to the WHO [World Health Organization] cancer is the second leading cause of death worldwide, and it is estimated that the number of new cases will increase in the coming years. There is enough evidence [Gottesman *et al.*, 2016; Bizzarri, *et al.*, 2011] regarding the complexity of cancer. We mean that cancer can be seen as a self-organizing nonlinear dynamic system, far from thermodynamic equilibrium [Deisboeck *et al.*, 2001; Rockmore, 2005; Izquierdo-Kulich & Nieto-Villar, 2013], exhibits a fractal structure that allows it to evade the action of the immune system and host cells [Bru, *et al.*, 2003; Kitano, 2004; Llanos-Perez *et al.*, 2015; Betancourt-Mar *et al.*, 2017]. The growth dynamics exhibit a deterministic chaos type, which confers high robustness, poor long-term prognosis, adaptability, and learning capacity [Itik, S. P. Banks, 2010; El-Gohary, 2008; Llanos-Perez *et al.*, 2016].

As we have shown in previous works [Montero *et al.*, 2018; Guerra, *et al.*, 2018; Betancourt-Mar *et al.*, 2017; Llanos-Perez *et al.*, 2015], cancer can be seen as a complex network made up of cells that have lost their specialization and growth control, and that emerges through what we can call "biological phase transition" [Montero *et al.*, 2018]. Indeed, even subtle changes in some critical values may impair the self-organization process, leading to unexpected different states, exhibiting variable robustness and adaptability capability within the attractor landscape [Montero *et al.*, 2018].

Overall, the development of a primary tumor from a microscopic level (avascular growth) to a macroscopic level (vascular phase), and the subsequent appearance of metastases, is not simply the accumulation of malignant cells, but results from a nonlinear process involving true "biological phase transition" downstream critical bifurcations [Betancourt-Mar *et al.*, 2017;

Martin, *et al.*, 2017]. This dynamical behavior leads to self-organization away from thermodynamic equilibrium, providing the system with a high degree of robustness, complexity, and hierarchy [Montero *et al.*, 2018], which, in turn, enacts the creation of new information and learning ability.

Many of the cancer therapies such as chemotherapy and radiotherapy are not specific, they have marked side effects. As a fact, it is well known that only 60% of different types of cancers can be healed through with the conventional therapies, they are also accompanied by undesirable side effects [Schulz, 2005].

Ferroptosis is a type of iron-dependent programmed cell death, characterized by the accumulation of free radicals and reactive oxygen species (ROS), such as, lipid peroxides, radical superoxide, hydroxyl radical, hydrogen peroxide and so on; and is genetically and biochemically distinct from other forms of regulated cell death such as apoptosis [Dixon *et al.*, 2012].

Three essential hallmarks define ferroptosis [Dixon & Stockwell, 2019], namely: The loss of lipid peroxide repair capacity by the phospholipid hydroperoxidase GPX4, the availability of redox-active iron and oxidation of polyunsaturated fatty acid (PUFA) containing phospholipids.

An open question is whether any type of lethal lipid peroxidation is classified as ferroptosis or whether only certain types of lethal lipid peroxidation should be termed ferroptosis [Feng & Stockwell, 2018]. In fact, the way lipid peroxidation leads to ferroptosis remains an unsolved mystery [Feng & Stockwell, 2018]. On one hand, exist the evidence that ferroptosis processes are associate to the pathogenesis of several degenerative diseases such as, cardiovascular disorders, cancer, atherosclerosis, diabetes, Alzheimer dementia (SDAT), just to mention the most relevant [Hong-fa Yan *et al.*, 2021; Ames *et al.*, 1993]. It leads to progressive loss of physiological integrity, leading to impaired function and increased vulnerability to death [Stockwell, *et al.*, 2017; Liochev, 2013; Harman, 2006].

On the other hand, numerous studies have demonstrated the effectiveness of cancer-killing by inducing ferroptosis, which is mainly accomplished by elevating the intracellular ROS levels and inactivating the activity of GPX4 [Shen, *et al.*, 2018; Shimada, *et al.*, 2016; Kim, *et al.*, 2016].

To the best of our knowledge, only a few models related to ferroptois have been reported [Kagan, *et al.*, 2017; Agmon, *et al.*, 2018; Konstorum, 2020]. Kagan *et al.* [Kagan, et al., 2017] developing a continuum model for ferroptosis with a focus on biochemical cascades. While that model provides an excellent synthesis of specific processes involved, it excludes contributions of the lipid peroxidation processes involved in ferroptosis. Agmon *et al.* [Agmon, *et al.*, 2018] performed molecular dynamics simulations of membranes with compositions relevant to ferroptotic sensitivity, and showed how the biophysical properties of membranes are altered under ferroptotic-competent lipid compositions. Most recently, Konstorum [Konstorum, 2020] has develop a multistate discrete modeling approach to emphasize qualitative properties of signaling cascades relevant to ferroptosis. The discrete modeling approach allows to explore the relative importance of different drivers of ferroptosis using a wider range of data than would be available for a detailed kinetic model of the system. As far as the authors know, there is no model that connects the lipid peroxidation processes involved in ferroptosis with the growth of cancer cells.

The main goal of the present work is establishing a heuristic model that connects lipid peroxidation, the evolution of the carcinogenic cells in the avascular and vascular phases [Roose, *et al.*, 2007] and the ferroptosis process. The model contains three populations species: $r(t)$-lipid peroxides, $x(t)$-avascular tumor cells and $y(t)$-vascular tumor cells. The manuscript is organized as follows: in Section 2 we propose a network model for the ferroptosis avascular and vascular tumor growth. Section 3 focuses into the analysis of the mathematical model derived from the mechanism previously proposed, including quantitative simulations and stability analyze. The development of a thermodynamic framework based on the entropy production rate is presented in Section 4. Finally, some concluding remarks are presented.

## 2. A network model of the ferroptosis

There is enough evidence and consensus in the literature related to ferroptosis-mediated anticancer effects [Wang, *et al.*, 2019; Dierge *et al.*, 2021; Hassannia, *et al.*, 2019]. However, the mechanisms underlying each step of this complex process remains unclear [Xu, *et al.*, 2019; Agmon, *et al.*, 2018; Feng & Stockwell, 2018]. Fig. 1 shows the network structure of the model proposed by us, where a connection is established between lipid peroxidation, the evolution of cancer in the avascular and vascular phases, respectively, and the ferroptosis processes.

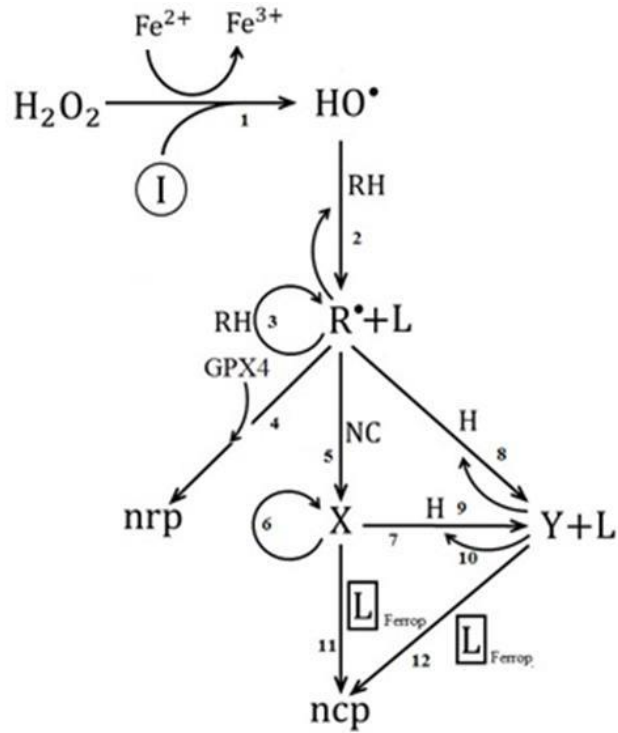

Fig. 1 The network model of ferroptosis avascular and vascular tumor growth.

In the network model, the symbol I represents the ferroptosis-inducing agents, such as iron-based nanomaterials, for example ferumoxytol, amorphous iron nanoparticles iron-organic frameworks, ionizing radiation-induced, and so on [Wang, et al., 2019; Zhang, et al., 2020]; $H_2O_2$ is a hydrogen peroxide; $HO^*$ are the hydroxyl radicals; RH represents the polyunsaturated fatty acids (PUFAs); NC are the population of normal cells exposed to the pro-carcinogenic stimulus; H the population of the host cells in the surrounding environment [Brú, et al., 2003], comprising exclusively epithelial cells; GPX4 glutathione peroxidase 4; L represents the population of the oxidized PUFAs fragments [Agmon, et al., 2018].

In this sense, we conjecture that L encourages the ferroptosis of cancer cells. For this reason we take $[L]_{CP}$ as the control parameter (CP).

Variables species: $r$, $x$, $y$ represent: $r(t)$ the population of lipid peroxides species, $x(t)$, $y(t)$ the population of tumor cells in the avascular and vascular state, respectively.. Finally, *nrp* and *ncp* represents a non-radical products and non-cancerous products respectively.

Step 1 is associated with the Fenton reaction [Cheng & Li, 2007], step 2 is related to formation of the Lipid peroxides [Bebber, 2020], step 3 is the propagate lipid peroxidation chain reactions [Spiteller & Afzal, 2014], step 4 is the primary cellular mechanism of protection against reactive oxygen species ( ROS ), which is mediated by glutathione peroxidase 4 (GPX4) [Dreger, et al., 2009], steps 5, 6 and 7, 10 are related to the process of mitosis and apoptosis of the proliferating tumor cells respectively; steps 8 and 9 correspond to the action of the host H [Brú, et al., 2003], finally, steps 11 and 12 are related to the ferroptosis avascular and vascular tumor growth.

The constants and concentration of species for the model proposed (see Fig. 1) were chosen empirically, dimensionless and trying to have a greater generality and simplicity as possible, so we have: $k_1 = 1$, $k_2 = 4$, $k_3 = 2$, $k_4 = 0.1$, $k_5 = 0.5$, $k_6 = 4$, $k_7 = 0.1$, $k_8 = 1$, $k_9 = 0.5$, $k_{10} = 0.001$, $k_{11} = k_{12} = 1$. $[RH] = 5$, $[GPX4] = 0.01$, $[H] = 3$, $[HO^*] = 0.001$, $[NC] = 1$, $[L]_{CP} = [4 - 0.3]$.

## 3. Mathematical model, stability analysis and numerical simulations

Mathematical models represent a suitable way for formalizing the knowledge of living systems obtained through a Systems Biology approach [Montero et al., 2018]. Mathematical modeling of tumor growth makes possible the description of its most important regularities and it is useful in providing effective guidelines for cancer therapy, drug development, and clinical decision-making [Preziosi, 2003; Araujo & McElwain, 2004; Bellomo, et al., 2008].

Although the role of the ferroptosis-mediated anticancer effects is well documented in literature, [Dierge et al., 2021] there are just few reports dealing with the ferroptois dynamics [Kagan, et al., 2017; Agmon, et al., 2018; Konstorum, 2020]. Indeed, most of the computational dynamic of ferroptois focus on the genetic and biophysical changes associated.

The network model we propose (Fig. 1) is a qualitative representation of the ferroptosis avascular and vascular tumor growth, based on the experimental evidences already available. In agreement with that model (Fig. 1) and the law of mass action governing chemical kinetics, a system of ordinary differential equations ODEs (1) was obtained, which describes the ferroptosis avascular and vascular tumor growth:

$$\frac{dr}{dt} = (a_1 - a_2 y + a_3) r - (a_4 + a_5) r^2$$

$$\frac{dx}{dt} = a_6 r^2 + (k_6 - k_{11} L) x - 2 k_7 x^2 - a_7 xy + k_{10} y^2 \qquad (3.1)$$

$$\frac{dy}{dt} = -k_{12} L y + k_7 x^2 + a_7 xy - 2 k_{10} y^2$$

where,

$$a_1 = k_3 (\text{RH}), a_2 = k_8 (\text{H}), a_3 = k_2 (\text{RH})(\text{HO}^*), a_4 = 2 k_5 (\text{NC}), a_5 = 2 k_4 (\text{GPX4}),$$
$$a_6 = k_5 (\text{NC}), a_7 = k_9 (\text{H}).$$

Fixed points, stability analysis and bifurcations were calculated using the standard procedure [Andronov & Khaikin, 1949; Anishchenko, et al., 2003; Kuznetsov, 2013]. The control parameter (CP) were represented the population of the oxidized PUFA fragments $[L]_{CP}$ [Agmon, et al., 2018]. Sensitivity analysis were done [Varma & Morbidelli, 2005] and quantitative investigation of the behavior of the output variables as the parameters of the system change.

The Lyapunov exponents were calculated using the Wolf algorithm in Fortran language [Wolf, et al., 1985]. Lyapunov dimension $D_L$, Eq. (2) also known as Kaplan–Yorke dimension [Frederickson, et al., 1983], was evaluated across the spectrum of Lyapunov exponents $\lambda_j$ as:

$$D_L = j + \frac{\sum_{i=1}^{j} \lambda_i}{|\lambda_{j+1}|} \qquad (3.2)$$

where $j$ is the largest integer number for which $\lambda_1 + \lambda_2 + ... + \lambda_j \geq 0$.

For simulation the network model, COPASI v.4.22.170 software was used. However, numerical integration was performed on the system of ODEs Eq. (3.1) through implementation of Gear algorithm for stiff equations, in Fortran with double precision and tolerance of $10^{-8}$ [Gear, 1968]. In Table 1 we show the dynamical behavior of the proposed ODEs (3.1) for different values of the control parameter $[L]_{CP}$.

**Table 1.** Stability, and complexity for the system of ODEs (3.1) for different values of the control parameter $[L]_{CP} = [4-0.3]$.

| L | Eigenvalues of the Jacobian matrix $\varepsilon_i$ | Lyapunov exponents $\lambda_j$ | Lyapunov dimension $D_L$ |
|---|---|---|---|
| 4<br>$ss_s$<br>stable focus | $\varepsilon_{1,2} = -0.158 \pm 5.37i$,<br>$\varepsilon_3 = -8.050$ | $\lambda_1 = -0.157$,<br>$\lambda_2 = -0.159$,<br>$\lambda_3 = -8.048$ | 0 |
| 3<br>Limit cycle | $\varepsilon_{1,2} = 0.083 \pm 4.80i$,<br>$\varepsilon_3 = -6.719$ | $\lambda_1 = 0.00$,<br>$\lambda_2 = -0.15$,<br>$\lambda_3 = -6.626$ | 1 |
| 0.57<br>P2 (2-period) | $\varepsilon_{1,2} = 0.273 \pm 2.29i$,<br>$\varepsilon_3 = -2.761$ | $\lambda_1 = 0.00$,<br>$\lambda_2 = 0.00$,<br>$\lambda_3 = -1.957$ | 2 |
| 0.42<br>P4 (4-period) | $\varepsilon_{1,2} = 0.249 \pm 2.01i$,<br>$\varepsilon_3 = -2.439$ | $\lambda_1 = 0.00$,<br>$\lambda_2 = 0.00$,<br>$\lambda_3 = -1.718$ | 2 |
| 0.3<br>Shilnikov's chaos | $\varepsilon_{1,2} = 0.220 \pm 1.73i$,<br>$\varepsilon_3 = -2.156$ | $\lambda_1 = 0.044$,<br>$\lambda_2 = 0.00$,<br>$\lambda_3 = -1.587$ | 2.102 |

In Fig. 2, the dynamic behavior of the network model is shown. It is observed as for low values of the control parameter $(L = 0.3)$, the tumor cells exhibit "apparently random behavior" (remnant of Shilnikov's chaos), (see Table 1) [Shilnikov, et al., 2001] (Fig. 2 A, B, C) with a predominance of the population of vascular tumor cells $y(t)$ (green). This behavior has important biological implications. On the one hand, the high sensitivity of the system to initial conditions makes unfeasible long-term predictions, i.e. the end forecasts are improbable (poor prognosis). Furthermore, the system displays a high degree of complexity [Betancourt-Mar & Nieto-Villar, 2007; Kitano, 2003]. This implies cancer cells are resilient in respect to pharmacological treatment, thus leading to a low response rate [Kim, et al., 2015].

As can be seen (see Fig. 2D) the increase of the population of the oxidized PUFA fragments, take as the control parameter $L$, produces an inverse Feigenbaum, (a cascade of saddle-foci Shilnikov's bifurcations) scenario, which results in the stabilization of the dynamics and in a decrease complexity of the system.

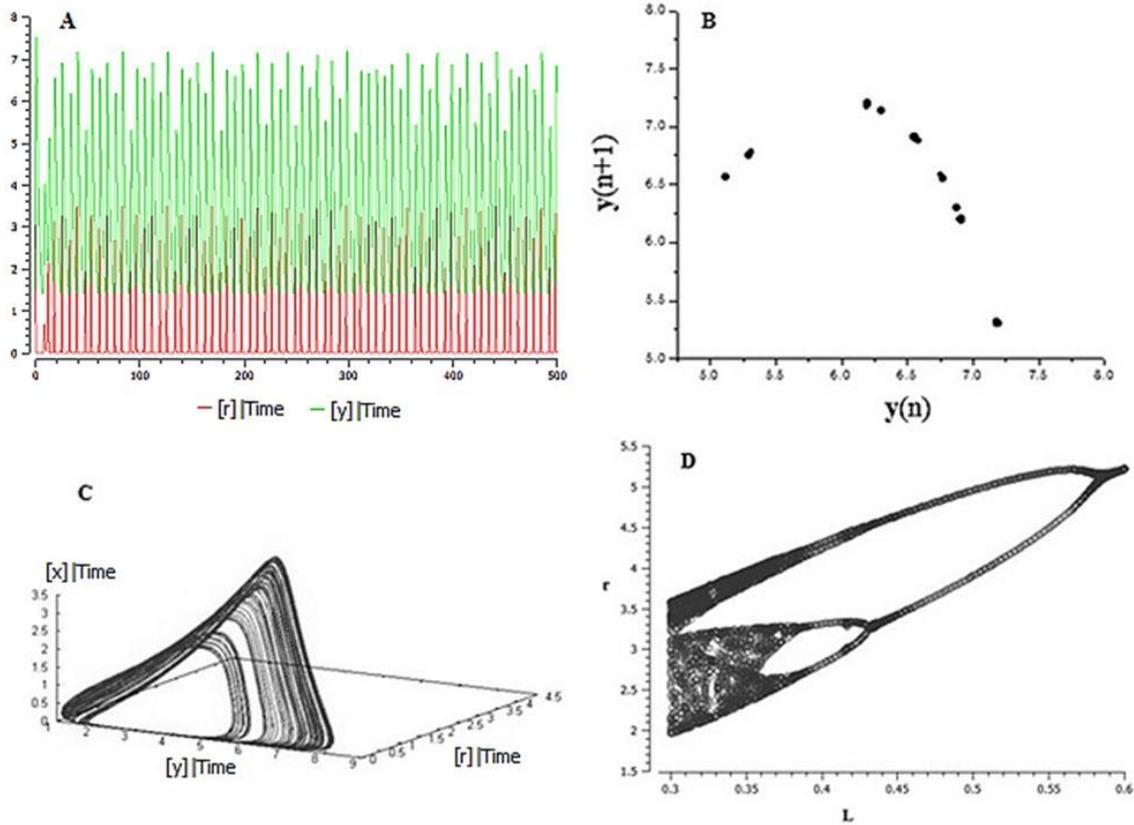

Fig. 2 Ferroptosis tumor growth dynamics for the proposed model (3.1), the control parameter values $(L=0.3)$: the population of lipid peroxides species $r$ (red) and $y$ (green) are the population vascular tumor cells; **A.** time series; **B.** Iterated unimodal map obtained from plotting successive local maxima of the time dynamics of vascular $y$ (green) tumor cells; **C.** Chaotic attractor; **D.** Bifurcation diagram obtained from $r_{max}$ showing the period-halving (i.e. inverse Feigenbaum) scenario occurring as the inactivation of the population of the oxidized PUFA fragments $L$ by tumor cells decreases. (For interpretation of the references to color in this figure legend, the reader is referred to the web version of this article).

As matter of the fact, at the critical point $L=3.383$, a supercritical Andronov–Hopf bifurcation takes place (Kuznetsov), and at $L=4$, giving rise to a stationary state. Furthermore, the sensitivity analysis of the model showed that the fundamental processes in decreasing order of their importance in the steady state $(L=4,\ ss_s)$ are the following: 3, 12, 11, 8 and 9. This shows how the propagation process of peroxidation lipid (3) and ferroptosis in avascular and vascular growth (12,11) have a pivotal position in our model.

Figure 3 shows the dynamic behavior for different values of the control parameter $L$.

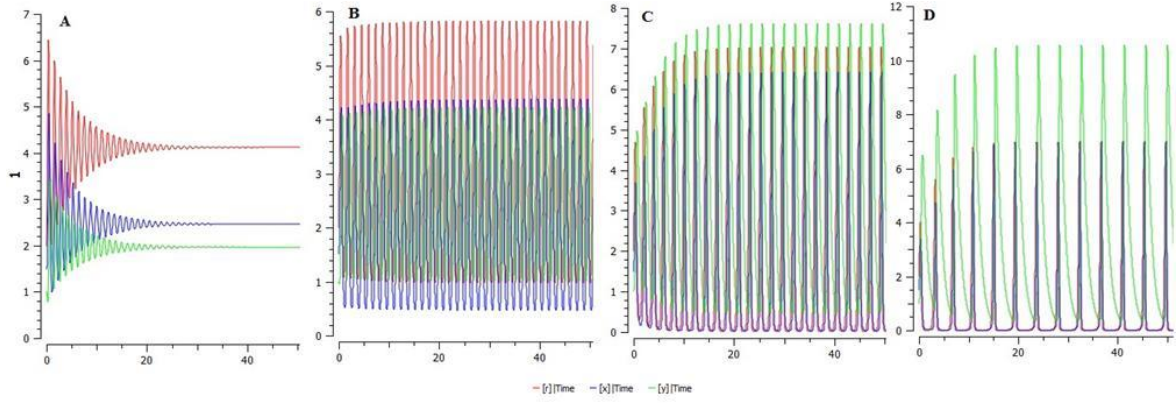

Fig. 3 Time series para diferentes valores del parametro de control $L$: **A**, steady state $(L = 4, \text{ ss}_s)$; Limit cycle (**B**, $L = 3$; **C**, $L = 2$; **D**, $L = 1$); $r$ (red), $x$ (blue) and $y$ (green) (For interpretation of the references to color in this figure legend, the reader is referred to the web version of this article).

In this way, we see that according to our conjecture, there is a fine regulation of the ferroptosis process of carcinogenic cells. In fact, as seen in Fig. 3, as the control parameter values decreases, there is a predominance of tumor cells compared to the other species and an increase in their population.

## 4. Thermodynamics framework

According to the seminal work of Posch–Hoover [Posch & Hoover, 1998] and Gaspard [Gaspard, 2004] the production of entropy per unit of time $\dfrac{dS_i}{dt}$ can be evaluate through of the spectrum of Lyapunov exponents $\lambda_j$ by means of the relationship

$$\frac{1}{k_B}\frac{dS_i}{dt} = -\sum_j \lambda_j > 0 \qquad (4.1)$$

where $\lambda_j$ are the spectrum of Lyapunov exponents.

In previous works we have shown how the entropy rate [Izquierdo-Kulich, et al., 2011] constitutes an index of robustness of tumors. As shown in Table 2, as the complexity (measured through the Lyapunov spectrum) of avascular and vascular tumor growth decreases, the ferroptosis process is favored, as we commented previously, which leads to an increase in its robustness $\left(\left.\dfrac{dS_i}{dt}\right|_{SS} = 8.37\right)$.

Table 2. Complexity vs. robustness for the ferroptosis of cancer cells

| Dynamic state | L | Complexity Lyapunov exponents $\lambda_j$ | Robustness $\dfrac{dS_i}{dt}$ |
|---|---|---|---|
| $ss_s$ stable focus | 4 | $\lambda_1 = -0.157$, $\lambda_2 = -0.159$, $\lambda_3 = -8.048$ | 8.37 |
| Limit cycle | 3 | $\lambda_1 = 0.00$, $\lambda_2 = -0.15$, $\lambda_3 = -6.626$ | 6.48 |
| P2 (2-period) | 0.57 | $\lambda_1 = 0.00$, $\lambda_2 = 0.00$, $\lambda_3 = -1.957$ | 1.98 |
| P4 (4-period) | 0.42 | $\lambda_1 = 0.00$, $\lambda_2 = 0.00$, $\lambda_3 = -1.718$ | 1.76 |
| Shilnikov's chaos | 0.3 | $\lambda_1 = 0.044$, $\lambda_2 = 0.00$, $\lambda_3 = -1.587$ | 1.55 |

## 5. Conclusions and remarks

Our model accurately captures the basic characteristics of the ferroptotic response, an important first step towards elucidate the main requirements of the ferroptosis process. In this sense, the proposed network model generalizes, at least qualitatively, the main features of the ferroptosis processes associates with avascular and vascular tumor growth. If the proposed conjecture (the PUFAs oxidized species are those that induce the ferroptosis process) is correct, the obtained bifurcation diagram can be used to establish different therapeutic strategies against cancer based on the stimulation of ferroptosis.

Summarizing, in this paper we have found that:

1. The ferroptosis appear as a type of "first order phase transitions", even for a range of discrete values of the control parameter $L$. Appraisal of ferroptosis as a process featured by criticality and threshold values may help in finding treatment strategies aimed to modify the overall process by targeting the singularities.

2. The avascular and vascular tumor growth exhibit Shilnikov's chaos dynamical behavior. The transition is tightly influenced by the control parameter $L$, representing by the represents the population of the oxidized PUFAs fragments.

We hope that the theoretical framework herewith described may help in establishing critical experiments that would improve our understanding of the ferroptosis processes in cancer evolution as well as finding optimal pathways for future treatments.

## Acknowledgements

Prof. Dr. Germinal Cocho and Prof. Dr. A. Alzola *in memoriam*. One of the authors (JMNV) thanked the CEIICH of the UNAM Mexico for the financial support.

## References

Agmon, E.; Solon, J.; Bassereau, P.; Stockwell, B.R. [2018] "Modeling the effects of lipid peroxidation during ferroptosis on membrane properties," Sci. Rep. 8, e5155.

Ames BN, Shigenaga MK, Hagen TM. [1993] "Oxidants, antioxidants, and the degenerative diseases of aging," Proceedings of the National Academy of Sciences 90(17):7915–7922.

Andronov AA, Khaikin SE. [1949] Theory of Oscillations (Princeton University Press).

Anishchenko VS, Vadivasova TE, Okrokvertskhov GA, Strelkova GI. [2003] "Correlation analysis of dynamical chaos," Physica A 325(1):199-212.

Araujo, R.P. and McElwain, D.L.S. [2004] "A history of the study of solid tumour growth: the contribution of mathematical modelling," Bull. Math. Biol. 66 (5) 1039–1091.

Bebber, Christina M. [2020] "Ferroptosis in Cancer Cell Biology," Cancers 12, 164; doi:10.3390/cancers12010164.

Bellomo, N., Li, N.K., Maini, P.K. [2008] "On the foundations of cancer modelling: selected topics, speculations, and perspectives," Math. Models Methods Appl. Sci. 18 (04) 593–646.

Betancourt-Mar J.A. & Nieto-Villar J.M. [2007] "Theoretical models for chronotherapy: periodic perturbations in funnel chaos type," Mathematical biosciences and engineering 4(2):177-86.

Betancourt-Mar, J.A. *et al.*, [2017] "Phase transitions in tumor growth: IV Relationship between metabolic rate and fractal dimension of human tumor cells," Physica A: Statistical Mechanics and its Applications, 473 344. https://doi.org/10.1016/j.physa.2016.12.089.

Bizzarri, M. *et al.*, [2011] "Fractal analysis in a systems biology approach to cancer," In Seminars in cancer biology 21 175. https://doi.org/10.1016/j. semcancer.2011.04.002.

Bru, A. *et al.*, [2003] "The universal dynamics of tumor growth," Biophysical journal, 85 2948. https://doi.org/10.1016/S0006-3495(03)74715-8.

Cheng, Z.; Li, Y. [2007] "What Is Responsible for the Initiating Chemistry of Iron-Mediated Lipid Peroxidation: An Update," Chem. Rev. 107, 748–766.

Deisboeck, T. S. *et al.*, [2001] "Pattern of self-organization in tumour systems: complex growth dynamics in a novel brain tumour spheroid model," Cell Prolif, 34 115. https://doi.org/10.1046/j.1365-2184.2001.00202.x.

Dierge E., *et al.*, [2021] "Peroxidation of n-3 and n-6 polyunsaturated fatty acids in the acidic tumor environment leads to ferroptosis-mediated anticancer effects," Cell Metabolism 33, 1–15.

Dixon, S.J. *et al.* [2012] "Ferroptosis: an iron-dependent form of nonapoptotic cell death" Cell, 149, 1060–1072.

Dreger, H. *et al.*, [2003] "Nrf2-dependent upregulation of antioxidative enzymes: Anovel pathway for proteasome inhibitor-mediated cardioprotection," Cardiovasc. Res. 83, 354–361.

El-Gohary, A. [2008] "Chaos and optimal control of cancer self-remission and tumor system steady states," Chaos Solitons Fractals 37 1305. https://doi.org/10.1016/j.chaos.2006.10.060.

Feng, H., and Stockwell, B.R. [2018] "Unsolved mysteries: How does lipid peroxidation cause ferroptosis?," PLoS Biol 16(5): e2006203. https://doi.org/10.1371/journal.pbio.2006203.

Frederickson, P., Kaplan, J.L., Yorke, E.D., Yorke, J.A. [1983] "The Liapunov dimension of strange attractors," J. Differential Equations 49 (2) 185–207.

Gaspard, P. [2004] "Time-reversed dynamical entropy and irreversibility in Markovian random processes," J. Stat. Phys. 117 599.

Gear, C.W. [1968] "The automatic integration of stiff ordinary differential equations," Proceedings IFIP68, (North-Holland, Amsterdam) pp. 187–193.

Guerra, A. *et al.* [2018] "Phase transitions in tumor growth VI: Epithelial-Mesenchymal transition," Physica A: Statistical Mechanics and its Applications. https://doi.org/10.1016/j.physa.2018.01.040.

Harman, D. [2006] "Free radical theory of aging: an update," Ann N Y Acad Sci. 1067(1):10–21.

Hassannia, B., Vandenabeele P. & Vanden Berghe T. [2019] "Targeting Ferroptosis to Iron Out Cancer," Cancer Cell 35, 830–849, doi:10.1016/j.ccell.2019.04.002.

Hong-fa Yan, *et al.*, [2021] "Ferroptosis: mechanisms and links with diseases," Signal Transduction and Targeted Therapy 6:49.

Itik, M. & Banks, S. P., [2010] "Chaos in a three-dimensional cancer model," Int. J. Bifurcation Chaos 20 71. https://doi.org/10.1142/S0218127410025417.

Izquierdo-Kulich, E., Alonso-Becerra E. & Nieto-Villar, J.M., [2011] "Entropy Production Rate for Avascular Tumor Growth," J. Modern Physics 2 615.

Izquierdo-Kulich, E. & Nieto-Villar, J.M., [2013] "Morphogenesis and Complexity of the Tumor Patterns," R.G. Rubio *et al.* (eds.), Without Bounds: A Scientific Canvas of Nonlinearity and Complex Dynamics. Understanding Complex Systems, (Springer-Verlag Berlin Heidelberg).

Kagan, V.E., *et al.*, [2017] "Oxidized arachidonic and adrenic PEs navigate cells to ferroptosis," Nat. Chem. Biol. 13 81–90. doi: 10.1038/nchembio.2238.

Kim, S.E. *et al.*, [2016] "Ultrasmall nanoparticles induce ferroptosis in nutrient-deprived cancer cells and suppress tumor growth," Nat. Nanotechnol. 11 977–985.

Kim, S., *et al.* [2015] "The Basic Helix-Loop-Helix Transcription Factor E47 Reprograms Human Pancreatic Cancer Cells to a Quiescent Acinar State With Reduced Tumorigenic Potential," Pancreas 44(5):718-27.

Kitano, H. [2003] "Cancer robustness: tumour tactics," Nature Publishing Group. 426:125.

Kitano, H. [2004] "Cancer as a robust system: implications for anticancer therapy," Nature 4 227. https://doi.org/10.1038/nrc1300.

34Konstorum, A., [2020] "Systems biology of ferroptosis: A modeling approach," Journal of Theoretical Biology 493 110222.

Kuznetsov, Y.A. [2013] Elements of Applied Bifurcation Theory (Springer Science & Business Media).

Lei, G., *et al.* [2020] "The role of ferroptosis in ionizing radiation-induced cell death and tumor suppression," Cell Res. 30:146–62.

Liochev, S.I. [2013] "Reactive oxygen species and the free radical theory of aging," Free Radic. Biol. Med. http://dx.doi.org/10.1016/j.freeradbiomed.2013.02.011i.

Llanos-Perez, J.A. *et al.*, [2015] "Phase transitions in tumor growth: II prostate cancer cell lines," Physica A: Statistical Mechanics and its Applications, 426 88. https://doi.org/10.1016/j.physa.2015.01.038.

39Llanos-Perez, J.A. *et al.*, [2016] "Phase transitions in tumor growth: III vascular and metastasis behavior," Physica A: Statistical Mechanics and its Applications 462 560. https://doi.org/10.1016/j.physa.2016.06.086.

Martin, R.R. *et al.* [2017] "Phase transitions in tumor growth: V what can be expected from cancer glycolytic oscillations?" Physica A: Statistical Mechanics and its Applications, https://doi.org/10.1016/j.physa.2017.06.001.

Montero, S. *et al.*, [2018] "Parameters Estimation in Phase-Space Landscape Reconstruction of Cell Fate: A Systems Biology Approach," In Systems Biology, 125-170, (Humana Press, New York, NY).

Posch, H.A., Hoover WG. [1988] "Lyapunov instability of dense Lennard-Jones fluids," Phys Rev A 38(1):473.

Preziosi, L. [2003] (Ed.), Cancer Modelling and Simulation, (CRC Press).

Rockmore, R. [2005] "Cancer complex nature," Santa Fe Institute Bulletin 20 18.

Roose, T., Chapman, S.J., Maini, P.K. [2007] "Mathematical models of avascular tumor growth," SIAM Rev. 49 (2) 179–208.

Schulz, W., [2005] Molecular biology of Human cancers (Springer Science).

Scott, J. Dixon and Brent, R. Stockwell, [2019] "The Hallmarks of Ferroptosis," Annu. Rev. Cancer Biol. 3:35–54.

Shen, Z., Song, J., Yung, B.C., [2018] "Emerging strategies of cancer therapy based on ferroptosis," Adv. Mater. 30 (12), 1704007.

Shilnikov, A.L., Turaev, D.V., Chua, L.O., [2001] Methods of Qualitative Theory in Nonlinear Dynamics, vol. 5, (World Scientific, Singapore).

Shimada, K. *et al.*, [2016] "Global survey of cell death mechanisms reveals metabolic regulation of ferroptosis," Nat. Chem. Biol. 12 (7), 497–503.

Shuaifei, W., *et al.*, [2019] "A mini-review and perspective on ferroptosis-inducing strategies in cancer therapy," Chinese Chemical Letters 30 847–852.

Spiteller, G. and Afzal, M. [2014] "The action of peroxyl radicals, powerful deleterious reagents, explains why neither cholesterol nor saturated fatty acids cause atherogenesis and age-related diseases," Chemistry - A European Journal, vol. 20, no. 46, pp. 14928–14945.

Stockwell, B.R. *et al.* [2017] "Ferroptosis: a regulated cell death nexus linking metabolism, redox biology, and disease," Cell 171, 273–285.

Tao. Xu, *et al.*, [2019] "Molecular mechanisms of ferroptosis and its role in cancer therapy," J Cell Mol Med. 23:4900–4912.

Varma, A., Morbidelli, M., Wu H., [2005] Parametric sensitivity in chemical systems (ed.; Cambridge University Press).

WHO (https://www.who.int/health-topics/cancer).

Wolf, A. *et al.*, [1985] "Determining Lyapunov exponents from a time series," Physica D 16 (3) 285–317.